\documentclass[runningheads]{llncs}
\usepackage{graphicx}
\usepackage{amssymb}
\usepackage{graphicx}
\usepackage{tikz}
\usetikzlibrary{arrows,calc,positioning}
\usetikzlibrary{automata}
\usepackage{pgfplots}
\pgfplotsset{compat=newest}
\usetikzlibrary{plotmarks}
\usepackage{grffile}
\usepackage{amsmath,amssymb}
\usetikzlibrary{shapes.gates.logic.US,shapes.gates.logic.IEC,calc}
\usepackage{url}
\usepackage{pdflscape}
\usepackage{afterpage}
\usepackage{wrapfig}
\usepackage{algorithm}
\usepackage{algpseudocode}
\usepackage{psfrag}
\usepackage{epstopdf}
\usepackage{cite}
\usepackage{caption}
\usepackage{subcaption}
\DeclareCaptionType{copyrightbox}
\usepackage{ifthen}
\usepackage{cleveref}
\usetikzlibrary{pgfplots.groupplots}
\usepackage{listings,lstautogobble}
\usepackage{color}
\usetikzlibrary{plotmarks,automata,arrows, positioning}
\usepackage{pgfplots}
\usepackage{adjustbox}
\usepackage{multirow}

\urldef{\mailsa}\path|{nathalie.cauchi, aabate}@cs.ox.ac.uk|  

 \newcommand{\eps}{\varepsilon}

 \newcommand{\HybridSpace}{\mathcal{H}}
 \newcommand{\normdis}{\mathcal{N}}
 
 \newcommand{\oC}{{ ~^\circ C}}

 \newcommand{\stochy}{$\mathsf{StocHy}$}
 \newcommand{\cpp}{\textsc{c++} }
 \newcommand{\shs}{\textsc{shs}}
  \newcommand{\pctl}{\textsc{pctl}}
   \newcommand{\ltl}{\textsc{ltl}}
   \newcommand{\bltl}{\textsc{bltl}}
   \newcommand{\csltl}{\textsc{csltl}}
 
 \newcommand{\matlab}{\textsc{matlab}}  
 \newcommand{\modest}{\textsc{modest toolset}}  
  \newcommand{\faust}{\textsc{faust}$^2$}    
   \newcommand{\imdp}{\textsc{imdp}}    
   \newcommand{\mdp}{\textsc{mdp}} 
    
    \newcommand{\python}{\textsc{python}}    
    \newcommand{\err}{\varepsilon}  
 \renewcommand{\phi}{\varphi}
\newcommand{\Id}{\mathbf{I}}

\newcommand{\Rl}{\mathbb{R}}     
\newcommand{\Na}{\mathbb{N}}     

\newcommand{\Q}{\mathcal{Q}}
\newcommand{\Sp}{\mathcal{D}}
\newcommand{\U }{\mathcal{U}}
\newcommand{\N }{\mathcal{N}}

\definecolor{lightblue}{RGB}{210,210,225}
\definecolor{lightred}{RGB}{235,220,220}
\definecolor{lightgreen}{RGB}{215,245,215}
\definecolor{lightyellow}{RGB}{225,222,200}
\definecolor{lightpurple}{RGB}{225,210,225}
\definecolor{darkblue}{RGB}{0,0,128}
\definecolor{darkred}{RGB}{128,0,0}
\definecolor{darkerred}{RGB}{64,0,0}
\definecolor{darkgreen}{RGB}{0,108,0}
\definecolor{darkyellow}{RGB}{25,22,0}
\definecolor{darkpurple}{RGB}{128,0,128}

\urldef{\mailsa}\path|{nathalie.cauchi, aabate}@cs.ox.ac.uk|  

\makeatletter
\renewcommand{\paragraph}{\@startsection{paragraph}{4}{0pt}%
	{.8ex plus 0.2ex minus 0.2ex}%
	{-0.5em}%
	{\bfseries}}
\makeatother

\newcommand{\colorpar}[3]{\colorbox{#1}{\parbox{#2}{#3}}}
\newcommand{\marginremark}[4]{
	\marginpar{\colorpar{#2}{\linewidth}{\color{#1}\tiny{[#3] #4}}}}

\makeatletter
\def\THICKhrulefill{\leavevmode \leaders \hrule height 5pt\hfill \kern \z@}
\makeatother

\iftrue
\newcommand{\rmkNMC}[1]{\marginremark{red}{red!25}{NMC}{#1}}
\newcommand{\rmkAA} [1]{\marginremark{darkpurple}{lightpurple}{AA}{#1}}

\else
\newcommand{\rmkNMC}[1]{}
\newcommand{\rmkAA}[1] {}
\fi

\newcommand{\nmc}[1]{\textcolor{black}{}}

\begin{document}
\author{%
	Nathalie Cauchi\inst{1}  \and 
	Kurt Degiorgio\inst{2} \and
	Alessandro Abate\inst{1}
}%
\institute{
	Department of Computer Science, University of Oxford, United Kingdom,
	\email{\{nathalie.cauchi, alessandro.abate\}@cs.ox.ac.uk},\\ 
	\and
	Diffblue Ltd, United Kingdom,\\
	\email{kurt.degiorgio@diffblue.com}
}   
\title{\stochy: automated verification and synthesis of stochastic processes }
\titlerunning{\stochy}
%
%
\authorrunning{Cauchi et al.}
%

%
\maketitle              
\begin{abstract}
\stochy~is a software tool for the quantitative analysis of {discrete-time} \textit{stochastic hybrid systems} (\shs).
\stochy~accepts a high-level description of stochastic models and {constructs} 
an equivalent \shs~model. 
The tool allows to 
(i) simulate the \shs~evolution over a given time horizon; 
and to automatically construct formal abstractions of the \shs. 
Abstractions are then employed for (ii) formal verification or (iii) control (policy, strategy) synthesis.  
\stochy~allows for modular modelling, 
and has separate simulation, verification and synthesis engines, 
which are implemented as independent libraries. 
This allows for libraries to be easily used and for extensions to be easily built.  
The tool is implemented in \cpp and employs manipulations based on vector calculus, 
the use of sparse matrices,  
the symbolic construction of probabilistic kernels, 
and multi-threading. 
Experiments show \stochy's markedly improved performance when compared to existing abstraction-based approaches: in particular, 
\stochy~beats state-of-the-art tools in terms of precision (abstraction error) and computational effort, 
and finally attains scalability to large-sized models (12 continuous dimensions).   
\stochy~is available at \url{www.gitlab.com/natchi92/StocHy}.
%
\end{abstract}
%
%
\section{Introduction}
\label{sec:intro}

\textit{Stochastic hybrid systems} (\shs) are a rich mathematical modelling framework capable of describing systems with complex dynamics, 
where uncertainty and hybrid (that is, both continuous and discrete) components are relevant. 
Whilst earlier instances of \shs~have a long history, 
\shs~proper have been thoroughly investigated only from the mid 2000s, 
and have been most recently applied to the study of complex systems, both engineered and natural.  
Amongst the first class, 
\shs~have been used for modelling and analysis of smart grids~\cite{Střelec2012_ISGT}, 
automation of medical devices~\cite{ARCH18}, 
avionics~\cite{BL06}, 
prognostics and health management~\cite{ZHAO201774}. 

However, a wider adoption of \shs~in real-world applications is stymied by a few factors: 
(i) the complexity associated with modelling \shs; 
(ii) the generality of their mathematical framework, 
which requires an arsenal of advanced and diverse techniques to analyse them; 
and (iii) the undecidability of verification/synthesis problems over \shs~and the curse of dimensionality associated with their approximations. 

This paper introduces a new software tool - \stochy - which is aimed at simplifying both the modelling of \shs~and their analysis,  
and targets the wider adoption of \shs~by non-expert users. 
With focus on the three limiting factors above, 
\stochy~allows to describe \shs~by parsing or extending well-known and -used state-space models  
and 
generates a standard \shs~model automatically and formats it to be analysed. 
\stochy~can 
(i) perform verification tasks, e.g., compute the probability of staying within a certain region of the state space from a given set of initial conditions;  
(ii) automatically synthesise policies (strategies) maximising this probability, 
and (iii) simulate the \shs~evolution over time. 
\stochy~is implemented in \cpp and modular making it both extendible and portable.

\paragraph{Related work.} 
There exist only a few tools that can handle (classes of) \shs. 
Of much inspiration for this contribution, 
\faust~\cite{soudjani2015faust} generates abstractions for uncountable-state discrete-time stochastic processes, 
natively supporting \shs~models with a single discrete mode and finite actions, 
and performs verification of reachability-like properties and corresponding synthesis of policies.  
\faust~is na\"ively implemented in \matlab~and lacks in scalability to large models. 
\modest~\cite{Hartmanns2014} allows to model and to analyse classes of continuous-time \shs, 
particularly probabilistic hybrid automata (\textsc{pha}) that combine probabilistic discrete transitions with deterministic evolution of the continuous variables. 
The tool for stochastic and dynamically coloured petri nets 
(\textsc{sdcpn})~\cite{everdij2006hybrid} supports compositional modelling of \textsc{pha} and focuses on simulation via Monte Carlo techniques. 
The existing tools highlight the need for a new software that allows for (i) straightforward and general \shs~modelling construction and (ii) scalable automated analysis. 

\paragraph{Contributions.} The \stochy~tool newly enables  
\begin{itemize}
	\item \textit{formal verification} of \shs~via either of two abstraction techniques:
	\begin{itemize}
		\item for discrete-time, continuous-space models with additive disturbances, and possibly with multiple discrete modes, 
		we employ formal abstractions as general Markov chains or Markov decision processes~\cite{soudjani2015faust}; \stochy\!\! 
		improves techniques in the \faust~tool by simplifying the input model description, by employing sparse matrices to manipulate the transition probabilities and by reducing the computational time needed  to generate the abstractions.
		\item for models with a finite number of actions, 
		we employ interval Markov decision processes and the model checking framework in~\cite{lahijanian2015formal}; 
		\stochy~provides a novel abstraction algorithm allowing for efficient computation of the abstract model, by means of an adaptive and sequential refining of the underlying abstraction. 
		We show that we are able to generate significantly smaller abstraction errors and to verify models with up to 12 continuous variables. 
	\end{itemize}
	\item \textit{control} (strategy, policy) \textit{synthesis} via formal abstractions, employing: 
	\begin{itemize}
		\item stochastic dynamic programming; \stochy~exploits the use of symbolic kernels.
		\item robust synthesis using interval Markov decision processes; \stochy~automates the synthesis algorithm with the abstraction procedure and the temporal property of interest, and exploits the use of sparse matrices;
	\end{itemize}
		\item \textit{simulation} of complex stochastic processes, such as \shs, by means of Monte Carlo techniques;  
		\stochy~automatically generates statistics from the simulations in the form of histograms, 
		visualising the evolution of both the continuous random variables and the discrete modes. 
\end{itemize}

This contribution is structured as follows: 
Sec.~\ref{sec:tf} crisply presents the theoretical underpinnings (modelling and analysis) for the tool.  
We provide an overview of the implementation of \stochy~in Sec.~\ref{sec:over}. 
We highlight features and use of \stochy~by a set of experimental evaluations in Sec.~\ref{sec:ev}:   
we provide four different case studies that highlight the applicability, ease of use, and scalability of \stochy. 
Details on executing all the case studies are detailed in this paper and within a Wiki page that accompanies the \stochy~distribution. 

\section{Theory: Models, Abstractions, Simulations}
\label{sec:tf}


\subsection{Models - Stochastic Hybrid Systems}

\stochy~supports the modelling of the following general class of \shs~\cite{APLS08,Abate2010:EUJournal_Control}.

\begin{definition}
A \shs~\cite{Abate2010:EUJournal_Control} is a discrete-time model defined as the tuple  
\begin{equation}\label{eqn:SHS:def}
\HybridSpace =(\Q,n,\U, T_{x}, T_{q}),  \quad where 
\end{equation}
\begin{itemize}
	\item $\Q$ = $ \{q_{1},q_{2}, \dots,q_{m}\}$, $m\in\Na,$ represents a finite set of modes (locations);
	\item $n\in\Na$ 
	is the dimension of the continuous space $\Rl ^{n}$ of each mode; 
	the hybrid state space is then given by $\Sp $= $\cup_{q\in \Q}\{q\} \times \Rl^{n} $; 
	\item $\U$ is a continuous set of actions, e.g. $\Rl^{v}$; 
	\item $T_{q}: \Q \times \Sp \times \U \rightarrow [0,1]$ is a discrete stochastic kernel on $Q$ given $\Sp \times \U$, which assigns to each $s=(q,x) \in \Sp$ and $u \in \U$, a probability distribution over $\Q: T_{q}(\cdot|s,u)$;
	\item $T_{x}: \mathcal B(\Rl^{n}) \times \Sp \times \U \rightarrow [0,1]$ is a Borel-measurable stochastic kernel on $\Rl^{n}$ given $\Sp\times \U$, 
	which assigns to each $s \in \Sp$ and $u \in \U$ a probability measure on the Borel space $(\Rl^{n}, \mathcal B(\Rl^{n})): T_{x}( \cdot |s,u)$.   
\end{itemize} 
\end{definition}
In this model the discrete component takes values in a finite set $\Q$ of modes (a.k.a. locations), 
each endowed with a continuous domain (the Euclidean space $\Rl^{n}$). 
As such, a point $s$ over the hybrid state space $\Sp$ is pair $(q, x)$, where $q \in \Q$ and $x \in \Rl^{n}$. 
The semantics of transitions at any point over a discrete time domain, are as follows: 
given a point $s \in \Sp$, 
the discrete state is chosen from $T_{q}$, 
and depending on the selected mode $q \in \Q$ the continuous state is updated according to the probabilistic law $T_{x}$.  
Non-determinism in the form of actions can affect both discrete and continuous transitions. 


\begin{remark}
A rigorous characterisation of \shs~can be found in \cite{APLS08}, 
which introduces a general class of models with probabilistic resets and a hybrid actions space. 
Whilst in principle we can deal with general \shs~models, 
in the case studies of this paper we focus on special instances, as described next. \qed
\end{remark}

\begin{remark}[Special instance]\label{rmk2}
In Case Study 2 (see Sec.~\ref{CS2}) we look at models where actions are associated to a deterministic selection of locations, namely $T_{q}: \U \rightarrow \Q$ and $\U$ is a finite set of actions.  
\qed 
\end{remark}

\begin{remark}[Special instance]
In Case Study 4 (Section \ref{CS4}) we consider non-linear dynamical models with bilinear terms, 
which are characterised for any $q \in \Q$ by $x_{k+1} = A_q x_{k}+ B_q u_k  + x_k\sum_{i=1}^{v}N_{q,i}u_{i,k}  + G_q w_k$, 
where $k \in \Na$ represents the discrete time index, 
$A_q,\; B_q,\;G_q$ are appropriately sized matrices, 
$N_{q,i}$ represents the bilinear influence of the $i-$th input component $u_{i}$, 
and
$w_k = w \sim \normdis(\cdot;0,1)$ and $\normdis(\cdot;\eta,\nu)$ denotes 
a Gaussian density function with mean $\eta$ and covariance matrix ${\nu^2}$.  
This expresses the continuous kernel $T_{x}: \mathcal B(\Rl^{n}) \times \Sp \times \U \rightarrow [0,1]$ as
\begin{align}
\normdis(\cdot; A_qx + B_qu + x\sum_{i=1}^{v}N_{q,i}u_{i}  + F_q, G_q). 
\label{eqn:ss}
\end{align} 
In Case Study~1-2-3 (Sec.~\ref{CS1}-\ref{CS3}), we look at the special instance from~\cite{lahijanian2015formal}, 
where the dynamics are autonomous (no actions) and linear: here $T_{x}$ is 
\begin{align}
\normdis(\cdot; A_qx+ F_q, G_q),   
\label{eqn:ss1}
\end{align} 
where in Case Studies~1, 3 $\Q$ is a single element.  
\qed
\end{remark}

\begin{definition}
A Markov decision process (\mdp)~\cite{baier2008principles} is a discrete-time model defined as the tuple  
\begin{equation}
\HybridSpace =(\Q,\U,T_{q}),  \quad where 
\end{equation}
\begin{itemize}
	\item $\Q$ = $ \{q_{1},q_{2}, \dots,q_{m}\}$, $m\in\Na,$ represents a finite set of modes; 
	\item $\U$ is a finite set of actions;  
	\item $T_{q}: \Q \times \Q \times \U \rightarrow [0,1]$ is a discrete stochastic kernel that 
	assigns, to each $q \in \Q$ and $u \in \U$, a probability distribution over $\Q: T_{q}(\cdot|q,u)$. 
\end{itemize}	
\label{def:mdp}
\end{definition}
Whenever the set of actions is trivial or a policy is synthesised and used (cf. discussion in Sec.~\ref{subsec:abstractions})   
the \mdp~reduces to a Markov chain (\textsc{mc}), 
and a kernel $T_{q}: \Q \times \Q \rightarrow [0,1]$ assigns to each $q \in \Q$ a distribution over $\Q$ as $T_q(\cdot|q)$. 

\begin{definition}
	An interval Markov decision process (\imdp)~\cite{vskulj2009discrete} extends the syntax of an \mdp~by allowing for uncertain $T_{q}$,  
	and is defined as the tuple 
	\begin{equation}
	\HybridSpace =(\Q,\U,\check{P},\hat{P}),  \quad where 
	\end{equation}
	\begin{itemize}
		\item $\Q$ and $\U$ are as in Def.~\ref{def:mdp};
		\item $\check{P}$ and $\hat{P}: \Q \times \U \times \Q \rightarrow [0,1]$ is a function that assigns to each $q \in \Q$ a \textit{lower} \textit{(upper)} bound probability distribution over $\Q: \check{P}(\cdot|q,u)$ $(\hat{P}(\cdot|q,u)$ respectively). 
	\end{itemize}	
	For all $q, q' \in \Q$ and $u \in \U$, it holds that $  \check{P}(q'|q,u) \leq \hat{P}(q' |q,u)$ and,
	$$ \sum_{q'\in \Q}  \check{P}(q' |q,u) \leq 1 \leq \sum_{q'\in \Q}  \hat{P}(q'|q,u).$$
	Note that when $\check{P}(\cdot |q,u) = \hat{P}(\cdot |q,u)$, the \imdp~reduces to the \mdp~with $\check{P}(\cdot |q,u) = \hat{P}(\cdot |q,u)= T_q(\cdot |q,u)$.
\end{definition}


\subsection{Formal Verification and Strategy Synthesis via Abstractions}
\label{subsec:abstractions}
Formal verification and strategy synthesis over \shs~are in general not decidable~\cite{Abate2010:EUJournal_Control,Summers20101951},  
and can be tackled via quantitative finite abstractions. 
These are precise approximations that come in two main different flavours: 
abstractions into \mdp~\cite{Abate2010:EUJournal_Control,soudjani2015faust} and into \imdp~\cite{lahijanian2015formal}. 
Once the finite abstractions are obtained, and with focus on specifications expressed in (non-nested) \pctl~or fragments of \ltl~\cite{baier2008principles}, 
formal verification or strategy synthesis can be performed via probabilistic model checking tools, 
such as \textsc{prism}~\cite{PRISM}, \textsc{storm}~\cite{storm}, \textsc{iscasMc}~\cite{iscasmc}.
We overview next the two alternative abstractions, as implemented in \stochy. 

\paragraph{Abstractions into Markov decision processes} 
Following \cite{soudjani2014formal}, \mdp~are generated by either 
(i) uniformly gridding the state space and computing an abstraction error, which depends on the continuity of the underlying continuous dynamics and on the chosen grid; 
or (ii) generating the grid adaptively and sequentially, by splitting the cells with the largest local abstraction error until a desired global abstraction error is achieved.  
The two approaches display an intuitive trade-off, where the first in general requires more memory but less time, whereas the second generates smaller abstractions.   
Either way, 
the probability to transit from each cell in the grid into any other cell characterises the \mdp~matrix $T_q$.  
Further details can be found in \cite{soudjani2015faust}. 
\stochy~newly provides a \cpp implementation and employs sparse matrix representation and manipulation,  
in order to attain faster generation of the abstraction and use in formal verification or strategy synthesis. 

\noindent\textit{Verification via \mdp} (when the action set is trivial) 
is performed to check the abstraction against 
non-nested, bounded-until specifications in \pctl~\cite{baier2008principles} or \textit{co-safe linear temporal logic} (\csltl) \cite{kupferman2001model}. 

\noindent\textit{Strategy synthesis via \mdp}
is defined as follows. Consider, the class of deterministic and memoryless Markov strategies $\pi = (\mu_0, \mu_1, \dots)$ where $\mu_k: \Q \rightarrow \U$. 
We compute the strategy $\pi^\star$ that maximises the probability of satisfying a formula, with 
algorithms discussed in~\cite{soudjani2015faust}. 

\paragraph{Abstraction into Interval Markov decision processes} (\imdp) 
is based on a procedure in~\cite{Cauchi2019HSCC}
performed using a uniform grid and with a finite set of actions $\U$ (see Remark~\ref{rmk2}). 
\stochy~newly provides the option to generate a grid using adaptive/sequential refinements (similar to the case in the paragraph above)~\cite{soudjani2014formal},  
which is performed as follows: 
(i) define a required minimal maximum abstraction error $\eps_{max}$; 
(ii) generate a coarse abstraction using the Algorithm in~\cite{Cauchi2019HSCC}  and compute the local error $\eps_q$ that is associated to each abstract state $q$; 
(iii) split all cells where $\eps_q > \eps_{max}$ along the main axis of each dimension, and update the probability bounds (and errors);  
and (iv) repeat this process until $\forall q, \; \eps_{q} < \eps_{max}$.
%
%
 
\noindent\textit{Verification via \imdp}
is run over properties in \csltl~or bounded-LTL (\bltl)~form using the 
\imdp~model checking algorithm in~\cite{lahijanian2015formal}. 

\noindent\textit{Synthesis via \imdp}~\cite{Cauchi2019HSCC}  
is carried out by extending the notions of strategies of \mdp~to depend on memory, that is on prefixes of paths.

\subsection{Analysis via Monte Carlo simulations}

Monte Carlo techniques generate numerical sampled trajectories representing the evaluation of a stochastic process over a predetermined time horizon. 
Given a sufficient number of trajectories, 
one can approximate the statistical properties of the solution process with a required confidence level. 
This approach has been adopted for simulation of different types of \shs. 
\cite{krystul2005sequential} applies sequential Monte Carlo simulation to \shs~to reason about rare-event probabilities. 
\cite{everdij2006hybrid} performs Monte Carlo simulations of classes of \shs~described as Petri nets. 
\cite{bouissou2014efficient} proposes a methodology for efficient Monte Carlo simulations of continuous-time \shs.  
In this work, we analyse a \shs~model using Monte Carlo simulations following the approach in~\cite{Abate2010:EUJournal_Control}. 
Additionally, we generate histogram plots at each time step, providing further insight on the evolution 
of the solution process. 

\section{Overview of \stochy}
\label{sec:over}

%


\paragraph{Installation}


\stochy~is set up using the provided \textsc{get\_dep} file found within the distribution package, which will automatically install all the required dependencies. 
The executable \textsc{run.sh} builds and runs \stochy. 
This basic installation setup has been successfully tested on machines running Ubuntu 18.04.1  LTS GNU and Linux operating systems.  

\begin{figure}[t]
	\begin{minipage}{\columnwidth}	
		\begin{lstlisting}[caption={Description of \textsc{main} file for simulating a \shs~consisting of two discrete modes and two continuous variables evolving according to~\eqref{eqn:ss}.},captionpos=b, label={lst:input},escapeinside=`']
		arma::mat Tq = { {0.4, 0.6},{0.7,0.3}};					// Transition probabilities
		// Evolution of the continuous variables for each discrete mode
		// First model
		arma::mat Aq0 = {{0.5, 0.4},{0.2,0.6}};
		arma::mat Fq0 = { {0},{0}};
		arma::mat Gq0 = {{0.4,0},{0.3, 0.3}};
		ssmodels_t modelq0(Aq0, Fq0, Gq0); 		
		// Second model
		arma::mat Aq1 = {{0.6, 0.3},{0.1,0.7}};
		arma::mat Fq1 = { {0},{0}};
		arma::mat Gq1 = {{0.2,0},{0.1, 0}};
		ssmodels_t modelq1(Aq1,Fq1, Gq1); 
		std::vector<ssmodels_t> models = 
		{modelq1,modelq2};
		// Set initial conditions	
		// Initial state q_0
		arma::mat q_init =  arma::zeros<arma::mat>(1,1); 					
		// Initial continuous variables
		arma::mat x1_init = arma::ones<arma:mat>(2,1);					
		exdata_t data(x1_init,q_init);
		// Build shs																
		shs_t<arma::mat,int> mySHS(Tq,models,data); 
		// Time horizon					 										
		int K = 32; 						
		// Task definition 	(1 = simulator, 2 = faust^2, 3 = imdp)
		int lb = 1;																																			 
		taskSpec_t mySpec(lb,K); 			
		// Combine								
		inputSpec_t<arma::mat,int> myInput(mySHS,mySpec);
		// Perform task										
		performTask(myInput);		
		\end{lstlisting}
	\end{minipage}
	\begin{minipage}{.5\columnwidth}
		\vspace{-20.2cm} \hspace{7cm}
		\includegraphics[width=.9\columnwidth]{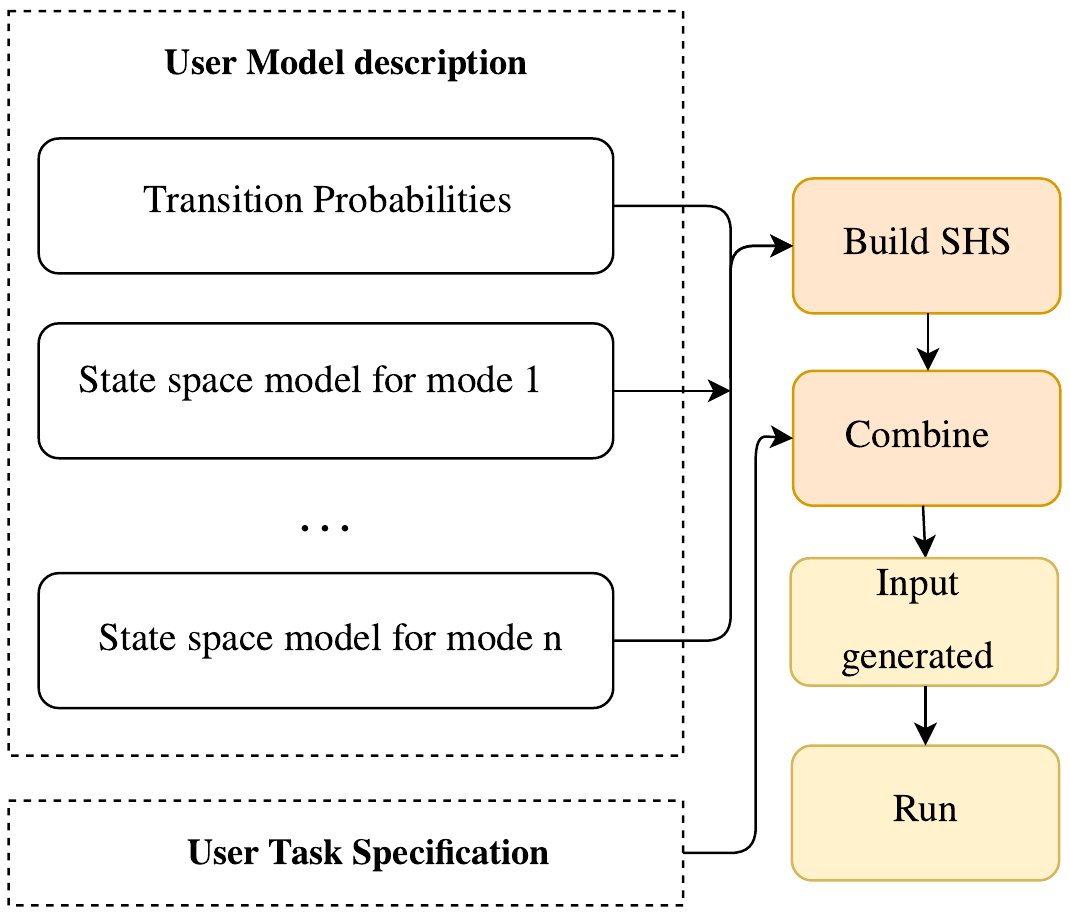}
	\end{minipage}
	\vspace{-1cm}
\end{figure}

\paragraph{Input interface}
The user interacts with \stochy~via the \textsc{main} file and must specify (i) a high-level description of the model dynamics and (ii) the task to be performed. 
The description of model dynamics can take the form of a list of the transition probabilities between the discrete modes, 
and of the state-space models for the continuous variables in each mode; 
alternatively, a description can be obtained by specifying a path to a \matlab~file containing the model description in state-space  form together with the transition probability matrix.  
Tasks can be of three kinds (each admitting specific parameters): simulation, verification, or synthesis. 
The general structure of the input interface is illustrated via an example in Listing~\ref{lst:input}:  
here the user is interested in simulating a \shs~with two discrete modes $\Q = \{q_0,q_1\}$ and  two continuous variables evolve according to~\eqref{eqn:ss1}. The model is autonomous and has no control actions.
The relationship between the discrete modes is defined by a fixed transition probability (line 1). 
The evolution of the continuous dynamics are defined in lines 2-14. 
The initial condition for both the discrete modes and the continuous variables are set in lines 16-21 (this is needed for simulation tasks).
The equivalent \shs~model is then set up by instantiating an object of type \texttt{shs\_t<arma::mat,int>} (line 23). 
Next, the task is defined in line 27 (simulation with a time horizon $K= 32$, as specified in line 25 and using the simulator library, as set in line 26). 
 We combine the model and task specification together in line 29. 
Finally, \stochy~carries out the simulation using the function \texttt{performTask} (line 31). 

\paragraph{Modularity}
\stochy~comprises independent libraries for different tasks, namely (i) \faust, (ii) \imdp, and (iii) simulator. 
Each of the libraries is separate and depends only on the model structure that has been entered. 
This allows for seamless extensions of individual sub-modules with new or existing tools and methods.  
The function \texttt{performTask} acts as multiplexer for calling any of the libraries depending on the input model and task specification.

\paragraph{Data structures}
\stochy~makes use of multiple techniques to minimise computational overhead. 
It employs vector algebra for efficient handling of linear operations, 
and whenever possible it stores and manipulates matrices as sparse structures. 
It uses the linear algebra library Armadillo~\cite{arma1,arma2}, 
which applies multi-threading and a sophisticated expression evaluator that has been shown to speed up matrix manipulations in \cpp when compared to other libraries. %
\faust~based abstractions define the underlying kernel functions symbolically using the library GiNaC~\cite{bauer2002introduction}, 
for easy evaluation of the stochastic kernels. 

\paragraph{Output interface}
We provide outputs as text files for all three libraries, which are stored within the \textsc{results} folder. 
We also provide additional \python~scripts for generating plots as needed. 
For abstractions based on \faust, the user has the additional option to export the generated \mdp~or \textsc{mc} to \textsc{prism} format, 
to interface with the popular model checker~\cite{PRISM} (\stochy~prompts the user this option following the completion of the verification or synthesis task). 
As a future extension, we plan to export the generated abstraction models to the model checker \textsc{storm}~\cite{storm} and to the modelling format \textsc{jani}~\cite{budde2017jani}.

\section{\stochy: Experimental Evaluation}
\label{sec:ev}

We apply \stochy~on four different case studies highlighting different models and tasks to be performed. 
All the experiments are run on a standard laptop, with an Intel Core i7-8550U CPU at 1.80GHz $\times$ 8 and with 8 GB of RAM. 

\subsection{Case Study 1 - Formal Verification}
\label{CS1}

We consider the \shs~model first presented in~\cite{Abate17memocode}. 
The model takes the form of~\eqref{eqn:SHS:def}, 
and has one discrete mode and two continuous variables representing the level of CO$_2$ ($x_1$) and the ambient temperature ($x_2$), respectively. 
The continuous variables evolve according to 
\begin{align}\label{eqn:CO2}
&x_{1,k+1} = x_{1,k} + \frac{\Delta}{V}(-\rho_mx_{1,k} + \varrho_{c}(C_{out} - x_{1,k})) + \sigma_{1} w_{k}, \\ 
&x_{2,k+1} = x_{2,k} + \frac{\Delta}{C_z}(\rho_mC_{pa} (T_{set} -x_{2,k}) +  \frac{\varrho_{c}}{R}(T_{out} - x_{2,k}))  + \sigma_{2} w_k, \nonumber 
\end{align}
where 
$\Delta$ the sampling time [$min$], 
$V$ is the volume of the zone [$m^3$],  
$\rho_m$ is the mass air flow pumped inside the room [$m^3/min$], 
$\varrho_{c}$ is the natural drift air flow  
[$m^3/min$], 
$C_{out}$ is the outside $CO_2$ level [$ppm/min$], 
$T_{set}$ is the desired temperature [$^oC$], 
$T_{out}$ is the outside temperature [$\oC/min$], 
$C_z$ is the zone capacitance [$Jm^3/\oC$], 
$C_{pa}$ is the specific heat capacity of air [$J/\oC$], 
$R$ is the resistance to heat transfer [$\oC/J$], 
and $\sigma_{(\cdot)}$ is a variance term associated to the noise $w_k \sim \normdis(0,1)$. 

We are interested in verifying whether the continuous variables remain within the safe set $X_{safe} =[405, 540] \times [18,24]$   
over 45 minutes ($K = 3$). 
This property can be encoded as a \bltl~property, 
 $\phi_1 := \square^{\le K} X_{safe},$
where $\square$ is the ``\textit{always}'' temporal operator considered over a finite horizon.  
The semantics of \bltl~is defined over finite traces, 
denoted by $\zeta = \{\zeta_j\}_{j=0}^{K}$. 
A trace $\zeta$ satisfies $\phi_1$ if $\forall j \le K, \zeta_j \in X_{safe}$,  
and we quantify the probability that traces generated by the \shs~satisfy $\phi_1$. 

When tackled with the method based on \faust~that hinges on the computation of Lipschitz constants,  
this verification task is numerically tricky, 
in view of difference in dimensionality of the range of $x_1$ and $x_2$ within the safe set $X_{safe}$ and the variance associated with each dimension $G_{q_0}= [\begin{smallmatrix}
 \sigma_{1} & 0 \\
 0 & \sigma_{2} \\
\end{smallmatrix}] = [\begin{smallmatrix}
40.096 &0 \\
0 & 0.511\\
\end{smallmatrix}]$. 
In order to mitigate this, we rescale the state space so all the dynamics evolve in a comparable range and also apply the abstraction based on \imdp. 
More precisely (this is done externally to \stochy), we consider an affine map $x = Jy$  with $J  = [\begin{smallmatrix}
22.5  & 0 \\
0 & 1\\
\end{smallmatrix}]$, 
which results in the safe set 
$ X_{safe}$ to $[18, 24]^2$ and in $G_{q_0} = [\begin{smallmatrix}
1.782&0 \\
0 & 0.511\\
\end{smallmatrix}]$. Consequently, the generated cell partitions are more uniform, with finer partitioning along $x_2$.
The dynamics of the new state space are provided in the file \textsc{cs1.mat}. 

\paragraph{Implementation}
\stochy~
provides two verification methods, one based on \faust\! \!
and the second based on \imdp. 
We parse the model from file 
\textsc{cs1.mat} (see line 2 of Listings~\ref{lst:CO2:verf}(a) and \ref{lst:CO2:veri}(b), corresponding to the two methods). \textsc{cs1.mat} sets parameter values to \eqref{eqn:CO2} and uses a $\Delta =$ 15 [$min$].
As anticipated, we employ both techniques over the same model description:  

\begin{figure}[t]
	\centering 
	
	\caption*{\textit{Case study 1:} Listings explaining task specification for (a) \faust~and (b) \imdp}
	\begin{minipage}{0.45\columnwidth}
		\begin{lstlisting}[caption={(a) \faust},captionpos=b, label={lst:CO2:verf}]
		// Dynamics definition
		shs_t<arma::mat,int> myShs('../CS1.mat');
		// Specification for FAUST^2
		// safe set
		arma::mat safe = {{18,24},{18,24}}; 
		// max error
		double eps = 1;
		// grid type 
		// (1 = uniform, 2 = adaptive)
		int gridType = 1;
		// time horizon
		int K = 3;
		// task and property type
		// (1 = verify safety , 2 = verify reach-avoid, 
		// 3 = safety synthesis, 4 = reach-avoid synthesis) 
		int p = 1; 
		// library (1 = simulator, 2 = faust^2, 3 = imdp)
		int lb = 2;	
		// task specification
		taskSpec_t mySpec(lb,K,p,safe,eps,gridType);
		\end{lstlisting}
	\end{minipage}
	\begin{minipage}{0.45\columnwidth}
		\begin{lstlisting}[caption={(b) \imdp},captionpos=b, label={lst:CO2:veri},numbers=none]
		// Dynamics definition
		shs_t<arma::mat,int> myShs('../CS1.mat');
		// Specification for IMDP
		// safe set
		arma::mat safe = {{18,24},{18,24}}; 
		// grid size for each dimension
		arma::mat grid = {{0.0845,0.0845}};
		// relative tolerance 
		arma::mat reft = {{1,1}};	
		// time horizon
		int K = 3;
		// task and property type
		// (1 = verify safety , 2 = verify reach-avoid, 
		// 3 = safety synthesis, 4 = reach-avoid synthesis) 
		int p = 1; 
		// library (1 = simulator, 2 = faust^2, 3 = imdp)
		int lb = 3;	
		// task specification
		taskSpec_t mySpec(lb,K,p,safe,grid,reft);
		\end{lstlisting}
	\end{minipage}\vspace{-.5cm}
\end{figure}
\begin{itemize}
\item 
for \faust~we specify the safe set ($X_{safe}$), the maximum allowable error, the grid type (whether uniform or adaptive grid), the time horizon, together with the type of property of interest (safety or reach-avoid). This is carried out in lines 5-21 in Listing~\ref{lst:CO2:verf}(a). 
\item 
for the \imdp~method, we define the safe set ($X_{safe}$), the grid size, the relative tolerance, the time horizon and the property type. 
This can be done by defining the task specification using lines 5-21 in Listing~\ref{lst:CO2:veri} (b).  
\end{itemize}

Finally, to run either of the methods on the defined input model, 
we combine the model and the task specification using \texttt{inputSpec\_t<arma::mat,int> myInput(myShs,mySpec)}, 
then run the command \texttt{performTask(myInput)}.  
The verification results for both methods are stored in the \textsc{results} directory: 
\begin{itemize}
\item 
for \faust, \stochy~generates four text files within the \textsc{results} folder: 
\textsc{representative\_points.txt} contains the partitioned state space;
\textsc{transition\_matrix.txt} consists of the transition probabilities of the generated abstract \textsc{mc}; 
\textsc{problem\_solution.txt} contains the sat probability for each state of the \textsc{mc};  
and \textsc{e.txt} stores the global maximum abstraction error. 
\item 
for \imdp, \stochy~generates three text files in the same folder: 
\textsc{stepsmin.txt} stores $\check{P}$ of the abstract \imdp; 
\textsc{stepsmax.txt} stores $\hat{P}$; 
and \textsc{solution.txt} contains the sat probability and the errors $\eps_q$ for each abstract state $q$. 
\end{itemize}
\begin{figure}[h]
	\vspace{-0.5cm}
	\begin{minipage}{.5\columnwidth}
		\centering
		\resizebox{\columnwidth}{!}{
			\begin{tabular}{ll|ccc}
				\textbf{Tool} & \textbf{Impl.} & $|{\mathbf{\Q}}|$    & \textbf{Time}  & \textbf{Error}\\
				\textbf{Method} & \textbf{Platform} & [states] & [s]           &   $\err_{\max}$\\ \hline \hline 
				\faust 	  & \matlab    & 576    & 186.746         &  1 \\
				\faust    & \cpp       & 576    & 51.420&  1 \\ 		
				\imdp     & \cpp       & 576    &  87.430 & 0.236\\
				\hline
				\faust 	  & \matlab    & 1089     &629.037 &  1 \\ 
				\faust    & \cpp       & 1089     & 78.140  &  1\\
				\imdp     & \cpp       & 1089     & 387.940  & 0.174 \\
				\hline
				\faust 	  & \matlab    & 2304    & 2633.155 &  1\\ 
				\faust    & \cpp       & 2304   &165.811  &  1\\ 
				\imdp     & \cpp       & 2304   &1552.950  & 0.121\\
				\hline
				\faust 	  & \matlab    & 3481   & 7523.771  &  1\\ 
				\faust    & \cpp       & 3481    & 946.294  &1 \\ 
				\imdp     & \cpp       & 3481     &  3623.090 & 0.098  \\ 
				\hline
				\faust 	  & \matlab    & 4225     &10022.850    		&0.900 \\ 
				\faust    & \cpp       & 4225    & 3313.990  & 0.900 \\  
				\imdp     & \cpp       & 4225    & 4854.580& 0.089 \\  
				\hline
		\end{tabular} }
		\captionof{table}{\textit{Case study 1:} Comparison of verification results for $\phi_1$ when using \faust~vs \imdp. }
		\label{tab:comp_FI}
	\end{minipage}
	~~~
	\begin{minipage}{0.45\columnwidth}
		\centering	
		\includegraphics[width=\columnwidth]{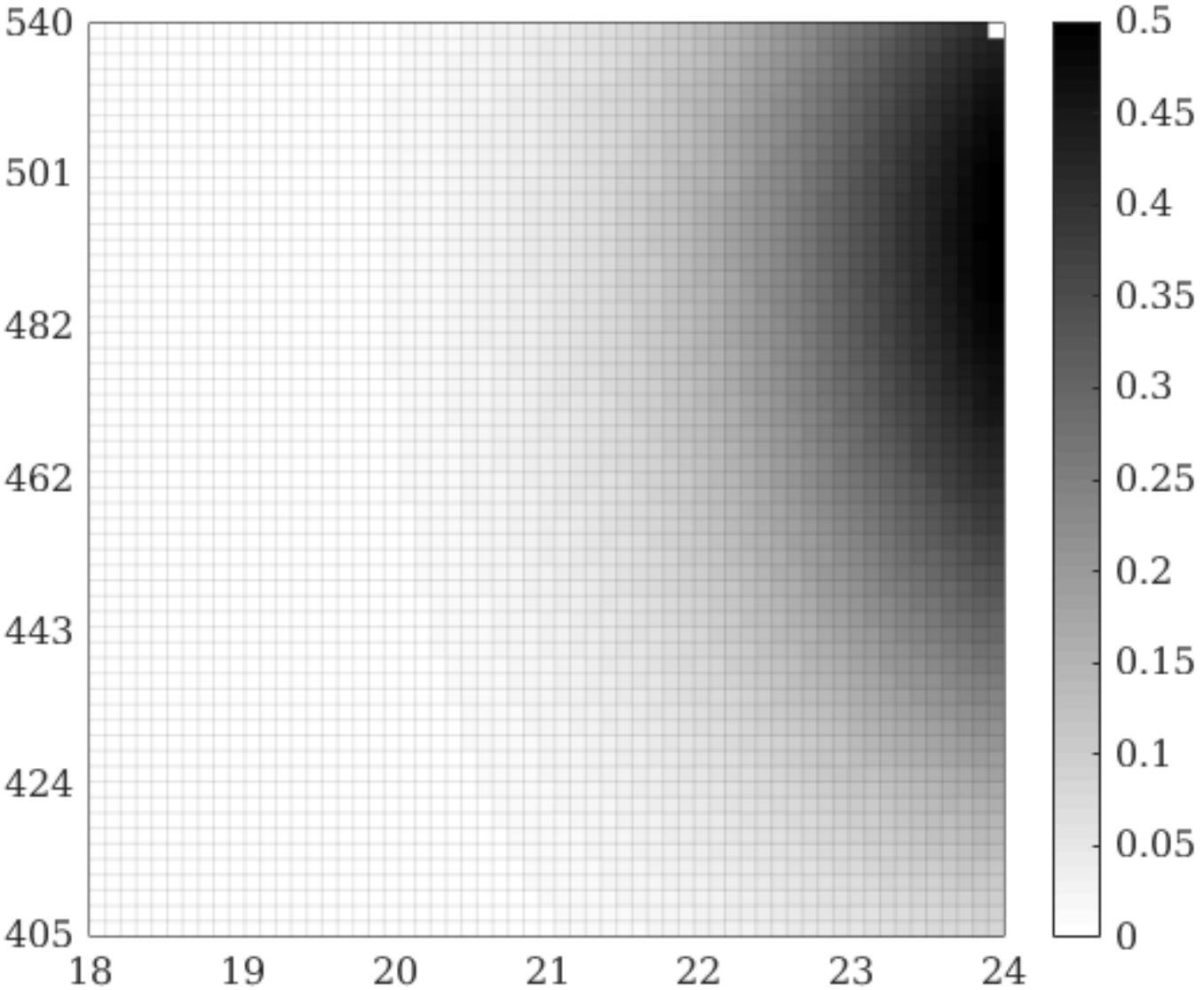}
		\vspace{4ex}
		\captionof{figure}{\textit{Case study 1: } Lower bound probability of satisfying $\phi_1$ generated using \imdp~ with $3481$ states.}
		\label{fig:Cs1lpb}
	\end{minipage}
\vspace{-0.5cm}
\end{figure}
\paragraph{Outcomes}
We perform the verification task using both \faust~and \imdp, over different sizes of the  abstraction grid. We employ uniform gridding for both methods. 
We further compare the outcomes of \stochy~against those of the \faust~tool, which is implemented in \matlab~\cite{soudjani2015faust}. 
Note that the \imdp~consists of $|\Q| + 1$ states, where the additional state is the sink state $q_u = \Sp \backslash X_{safe}$.
The results are shown in Table~\ref{tab:comp_FI}.  
We saturate (conservative) errors output that are greater than $1$ to this value. 
We show the probability of satisfying the formula obtained from 
\imdp~for a grid size of 3481 states in Fig.~\ref{fig:Cs1lpb} -- similar probabilities are obtained for the remaining grid sizes. As evident from Table~\ref{tab:comp_FI}, the new \imdp~method outperforms the approach using \faust~in terms of the maximum error associated to the abstraction (\faust~generates an abstraction error $< 1$ only with 4225 states).   
Comparing the \faust~within \stochy~and the original \faust~implementation (running in \matlab), \stochy~offers  computational speed-up for the same grid size. This is due to the faster computation of the transition probabilities, through \stochy's use of matrix manipulations.
\faust~within \stochy~also simplifies the input of the dynamical model description: in the original \faust~implementation, the user is asked to manually input the  stochastic kernel in the form of symbolic equations in a \matlab~script. 
This is not required when using \stochy, which automatically generates the underlying symbolic kernels from the input state-space model descriptions.


\subsection{Case Study 2 - Strategy Synthesis }
\label{CS2}
We consider a stochastic process with two modes $\Q = \{q_0, q_1\} $, which continuously evolves according to~\eqref{eqn:ss1} with  
\begin{equation*}
A_{q_0} =  
\begin{bmatrix}
0.43& 0.52\\
0.65& 0.12\\
\end{bmatrix},
G_{q_0}= 
\begin{bmatrix}
1&0.1\\
0&0.1 \\
\end{bmatrix},
A_{q_0} =  
\begin{bmatrix}
0.65 & 0.12\\ 
0.52 & 0.43\\
\end{bmatrix},
G_{q_1} = 
\begin{bmatrix}
0.2 &0\\
0& 0.2\\
\end{bmatrix},
F_{q_{i}} = \begin{bmatrix}
0\\
0\\
\end{bmatrix},
\end{equation*}
\begin{flushright}
	\begin{figure}[t]
	\centering
	\begin{subfigure}{.3\columnwidth}
		\centering
		\includegraphics[width=.9\columnwidth]{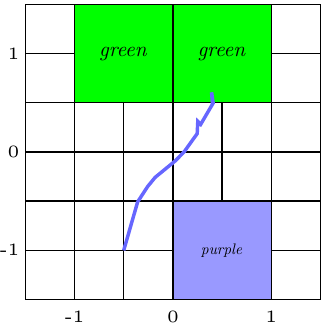}
		\caption{}
		\label{fig:Cs2:domain}
	\end{subfigure}
	\begin{subfigure}{.34\columnwidth}
		\centering
		\includegraphics[width=.94\columnwidth]{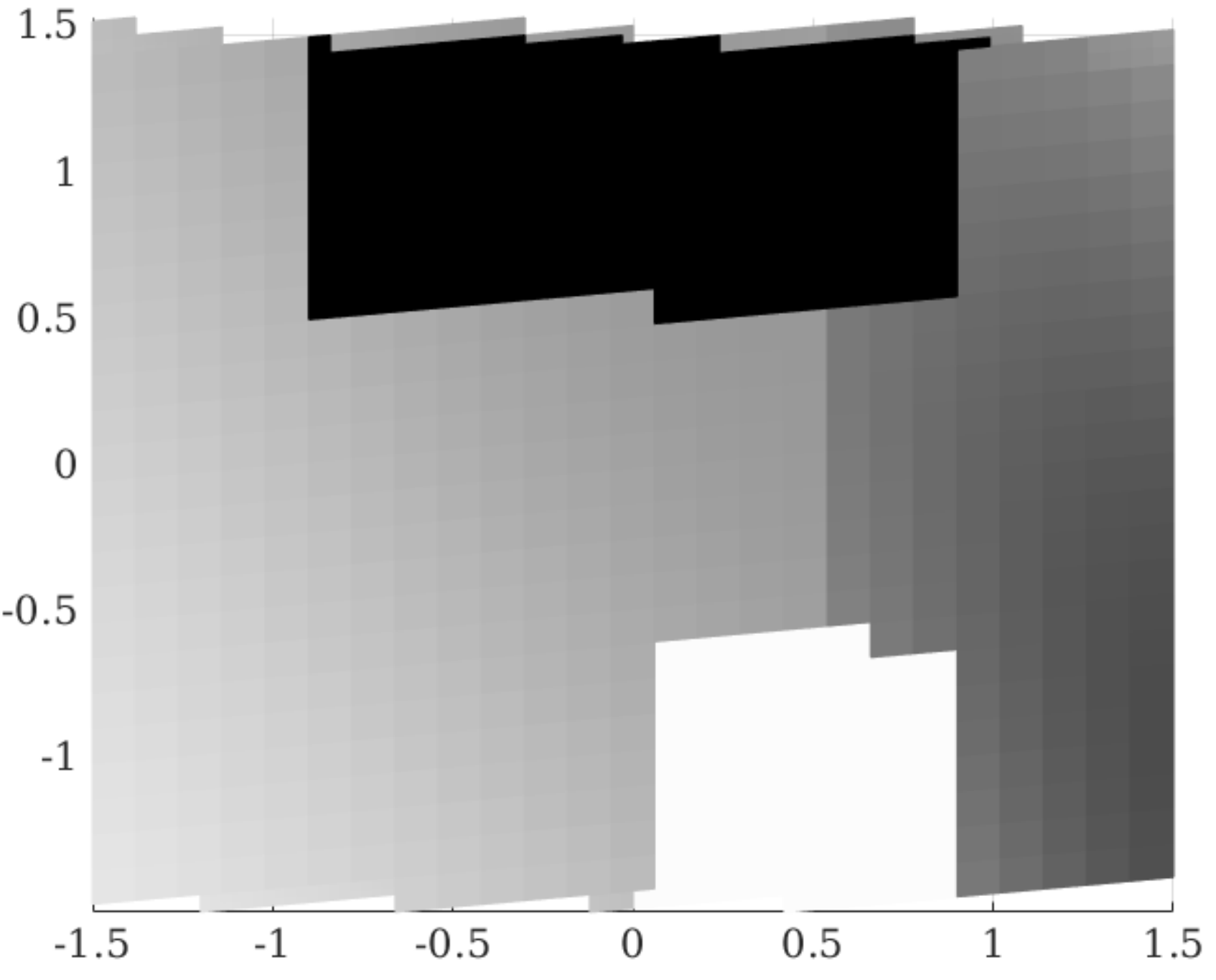}
		\caption{}
		\label{fig:Cs2:Lq1}
	\end{subfigure}
	\begin{subfigure}{.34\columnwidth}
		\centering
		\includegraphics[width=1.05\columnwidth]{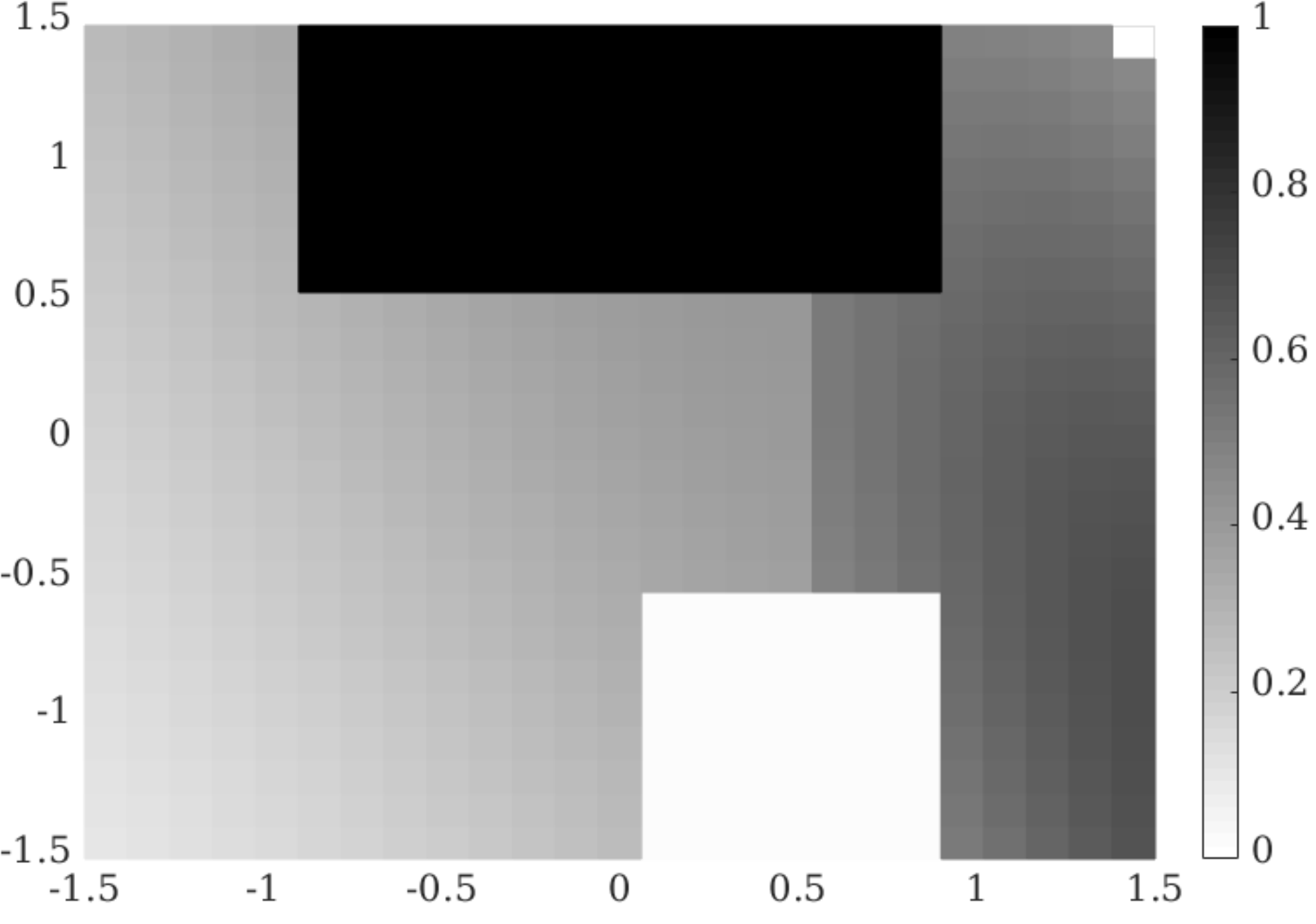}
		\caption{}
		\label{fig:Cs2:Lq2}
	\end{subfigure}
	\caption{ \textit{Case study 2:} (a) Gridded domain together with a superimposed simulation of trajectory initialised at $(-0.5,-1)$ within $q_0$, under the synthesised optimal switching strategy $\pi^*$. Lower probabilities of satisfying $\phi_2$ for mode $q_0$ (b) and for mode $q_1$ (c), as computed by \stochy. }%
	\vspace{-.3cm}
\end{figure}
\end{flushright}
and $i\in\{0,1\}$. 
Consider the continuous domain shown in Fig.\ref{fig:Cs2:domain} over both discrete locations. 
We plan to synthesise the optimal switching strategy $\pi^\star$ that maximises the probability of reaching the \textit{green} region, 
whilst avoiding the \textit{purple} one, 
over an unbounded time horizon, given any initial condition within the domain. 
This requirement can be expressed with the \ltl~formula,
$\phi_2 :=  (\neg purple) \;\mathsf{U}\;green,$ 
where $\mathsf{U}$ is the ``\textit{until}'' temporal operator, 
and the atomic propositions $\{purple,$ $ green\}$ 
denote regions within the set $X = [-1.5, 1.5]^2$,  
as shown in Fig.~\ref{fig:Cs2:domain}.

\paragraph{Implementation}
We define the model dynamics following lines 3-14 in Listing~\ref{lst:input}, while we use Listing~\ref{lst:CO2:veri} to specify the synthesis task and together with its associated parameters.  
The \ltl~property $\phi_2$
is over an unbounded time horizon, which leads to employing the \imdp~method for synthesis 
(recall that the \faust~implementation can only handle time-bounded properties, and its abstraction error monotonically increases with the time horizon of the formula).  
In order to encode the task we set the variable \texttt{safe} to correspond to $X$ the grid size to $0.12$ and the relative tolerance to $0.06$ along both dimensions (cf. lines 5-10 in Listing~\ref{lst:CO2:veri}). We set the time horizon \texttt{K = -1} to represent an unbounded time horizon,  let \texttt{p = 4} to trigger the synthesis engine over the given specification and make \texttt{lb = 3} to use \imdp~method (cf. lines 12-19 in Listing~\ref{lst:CO2:veri}).
This task specification
partitions the set $X$ into the underlying \imdp~via uniform gridding. Alternatively, the user has the option to make use of the adaptive-sequential algorithm by defining a new variable \texttt{eps\_max} which characterise the maximum allowable abstraction error and then specify the task using \texttt{taskSpec\_t mySpec(lb,K,p,boundary,eps\_max,grid,rtol);}.
Next, we define two files (\textsc{phi1} \textsc{.txt} and \textsc{phi2.txt}) containing the coordinates within the gridded domain (see Fig.\ref{fig:Cs2:domain}) associated with the atomic propositions \textit{purple} and \textit{green}, respectively. This allows for automatic labelling of the state-space over which synthesis is to be performed. 
Running the main file, \stochy~generates a \textsc{Solution.txt} file within the \textsc{results} folder. 
This contains the synthesised $\pi^\star$ policy, the lower bound for the probabilities of satisfying $\phi_2$, and the local errors $\eps_q$ for any region $q$. 

\paragraph{Outcomes} 
The case study generates an abstraction with a total of 2410 states, a maximum probability of 1,
a maximum abstraction error of 0.21, and it requires a total time of 1639.3 [s].
In this case, we witness a slightly larger abstraction error via the \imdp~method then in the previous case study.
This is due the non-diagonal covariance matrix $G_{q_0}$  which introduces a rotation in 
$X$ within mode $q_0$. 
When labelling the states associated with the regions  $purple$ and $green$, 
an additional error is introduced due to the over- and under-approximation of states associated with each of the two regions. 
We further show the simulation of a trajectory under  $\pi^\star$ with a starting point of $(-0.5,-1)$ in $q_0$, within Fig.\ref{fig:Cs2:domain}. 

\subsection{Case Study 3 - Scaling in Continuous Dimension of Model}
\label{CS3}

We now focus on the continuous dynamics by considering a stochastic process with $\Q =\{q_0\}$ (single mode) and dynamics evolving according to~\eqref{eqn:ss1}, characterised by $A_{q_0} =  0.8\Id_{d}$, $F_{q_0} = \textbf{0}_d$  and $G_{q_0}= 0.2\Id_{d}$, 
where $d$ corresponds to the 
number of continuous variables. 
We are interested in checking the \ltl~specification 
$\phi_3 := \square X_{safe}$, 
where $X_{safe}= [-1,1]^{d}$, as the continuous dimension $d$ of the model varies. Here ``$\square$'' is the ``\textit{always}'' temporal operator and 
a trace $\zeta$ satisfies $\phi_3$ if $\forall k \ge 0,\; \zeta_k \in X_{safe}$.  
In view of the focus on scalability for this Case Study 3, we disregard discussing the computed probabilities, which we instead covered in Section \ref{CS1}. 

\paragraph{Implementation}
Similar to Case Study~2, we follow lines 3-14 in Listing~\ref{lst:input} to define the model dynamics, while we use Listing~\ref{lst:CO2:veri} to specify the verification task using 
 the \imdp~method. 
For this example, we employ a uniform grid having a grid size of $1$ and relative tolerance of $1$ for each dimension (cf.  lines 5-10 in Listing~\ref{lst:CO2:veri}).
We set \texttt{K = -1} to represent an unbounded time horizon, \texttt{p = 1} to perform verification over a safety property and  \texttt{lb = 3} to use the \imdp~method (cf. lines 12-19 in Listing~\ref{lst:CO2:veri}).
In Table~\ref{tab:comp_Dim} 
we list the number of states required for each dimension, 
the total computational time, 
and the maximum error associated with each abstraction.

\begin{table}[t]
\vspace{-0.3cm}
	\centering	
	\resizebox{\columnwidth}{!}{
	\begin{tabular}{c||c|c|c|c|c|c|c|c|c|c|c}  \hline
	  \textbf{Dimensions} & \multirow{2}{*}{2}& \multirow{2}{*}{3} &\multirow{2}{*}{4} & \multirow{2}{*}{5} & \multirow{2}{*}{6}& \multirow{2}{*}{7} & \multirow{2}{*}{8} & \multirow{2}{*}{9} & \multirow{2}{*}{10} & \multirow{2}{*}{11} & \multirow{2}{*}{12}  \\   
	  $[\text{d}]$    & & & & & & & & & & &  \\ \hline 
	  
	 \textbf{$|\mathbf{\Q}|$} & \multirow{2}{*}{4}& \multirow{2}{*}{14} &\multirow{2}{*}{30} & \multirow{2}{*}{62} & \multirow{2}{*}{126}& \multirow{2}{*}{254} & \multirow{2}{*}{510} & \multirow{2}{*}{1022} & \multirow{2}{*}{2046} & \multirow{2}{*}{4094} & \multirow{2}{*}{8190}  \\   
	 \text{ [states]}   		& & & & & & & & & & &   \\ \hline 

	 \textbf{Time taken} & \multirow{2}{*}{0.004 }& \multirow{2}{*}{0.06} &\multirow{2}{*}{0.21 } & \multirow{2}{*}{0.90 } & \multirow{2}{*}{ 4.16}& \multirow{2}{*}{19.08} & \multirow{2}{*}{79.63 } & \multirow{2}{*}{319.25} & \multirow{2}{*}{1601.31} & \multirow{2}{*}{5705.47} & \multirow{2}{*}{21134.23}  \\   
	  \text{ [s]}   		& & & & & & & & & &  &   \\ \hline 
	  
	 \textbf{Error} & \multirow{2}{*}{4.15e-5}& \multirow{2}{*}{3.34e-5} &\multirow{2}{*}{2.28e-5} & \multirow{2}{*}{ 9.70e-5 } & \multirow{2}{*}{8.81e-6}& \multirow{2}{*}{1.10e-6 } & \multirow{2}{*}{ 2.95e-6 } & \multirow{2}{*}{ 4.50e-7} & \multirow{2}{*}{1.06e-7} & \multirow{2}{*}{ 4.90e-8} & \multirow{2}{*}{4.89e-8 }  \\   
	$(\err_{max})$    		& & & & & & & & & &  &   \\ \hline 
	\end{tabular}}
\vspace{0.1cm}
	\caption{\textit{Case study 3:} 
		Verification results of the \imdp-based approach over $\phi_3$,  
		for varying dimension $d$ of the stochastic process. } 
	\label{tab:comp_Dim}
	\vspace{-1cm}
\end{table}

\paragraph{Outcomes} 
From Table~\ref{tab:comp_Dim} we can deduce that by employing the \imdp~method within \stochy, 
the generated abstract models have manageable state spaces, 
thanks to the tight error bounds that is obtained.  
Notice that since the number of cells per dimension is increased with the dimension $d$ of the model, 
the associated abstraction error $\err_{max}$ is decreased. 
The small error is also due to the underlying contractive dynamics of the process. 
This is a key fact leading to scalability over the continuous dimension $d$ of the model: 
\stochy~displays a significant improvement in scalability over the state of the art~\cite{soudjani2015faust} and allows abstracting stochastic models with relevant dimensionality.  
Furthermore, \stochy~is capable to handle specifications over infinite horizons (such as the considered \textit{until} formula). 

\vspace{-0.25cm}
\subsection{Case Study 4 - Simulations}
\label{CS4}

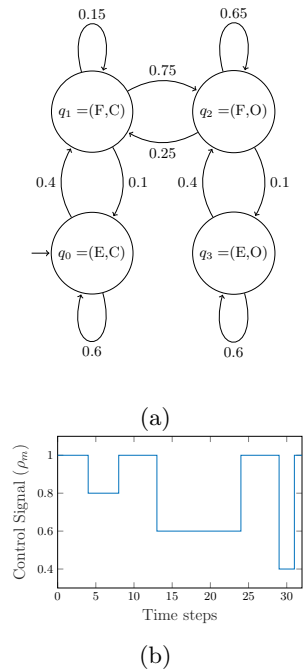
\begin{figure}[h!]
	\vspace{-0.1cm}
	\centering	
	\begin{minipage}{0.32\columnwidth}
		\begin{subfigure}{\columnwidth}
		\begin{center}
			\resizebox{.95\columnwidth}{!}{
				\begin{tikzpicture}[->,shorten >=1pt,initial text={},auto,node distance=3cm,
				transform shape]
				\tikzstyle{every state}=[fill=white,text=black]	
				\node[state] (A)              {$q_1 =$(F,C)};
				\node[state]         (B) [right of=A] {$q_2 =$(F,O) };
				\node[initial,state]         (E) [below of=A] {$q_0 =$(E,C) };
				\node[state]         (F) [below of=B] { $q_3 =$(E,O)};
				\path (B) edge  [bend left,sloped, anchor=center, below] node {$0.25$} (A)
				(A) edge [loop above] node {$0.15$} (A)
				(A) edge  [bend left] node {$0.1$} (E)
				(A) edge  [bend left] node {$0.75$} (B)
				(B) edge [loop above] node {$0.65$} (B)
				(E) edge  [bend left] node {$0.4$} (A)
				(F) edge  [bend left] node {$0.4$} (B)
				(E) edge  [loop below] node {$0.6 $} (E)
				(F) edge  [loop below] node {$0.6$} (F)
				(B) edge  [bend left] node {$ 0.1$} (F);	
				\end{tikzpicture}}	
		\end{center}
		\caption{}
		\label{fig:SHS:CO2}	
		\end{subfigure}
			\begin{subfigure}{\columnwidth}
			\centering
			\resizebox{\columnwidth}{!}{\definecolor{mycolor1}{rgb}{0.00000,0.44700,0.74100}%
\begin{tikzpicture}

\begin{axis}[%
width=2.521in,
height=1.566in,
scale only axis,
xmin=0,
xmax=32,
xlabel style={font=\color{white!15!black}},
xlabel={\large Time steps},
ymin=0.3,
ymax=1.1,
ylabel={\large Control Signal ($\rho_m$)},
ylabel style={font=\color{white!15!black}},
axis background/.style={fill=white},
legend style={legend cell align=left, align=left, draw=white!15!black}
]
\addplot[const plot, color=mycolor1] table[row sep=crcr] {%
0 1\\
1	1\\
2	1\\
3	1\\
4	0.8\\
5	0.8\\
6	0.8\\
7	0.8\\
8	1\\
9	1\\
10	1\\
11	1\\
12	1\\
13	0.6\\
14	0.6\\
15	0.6\\
16	0.6\\
17	0.6\\
18	0.6\\
19	0.6\\
20	0.6\\
21	0.6\\
22	0.6\\
23	0.6\\
24	1\\
25	1\\
26	1\\
27	1\\
28	1\\
29	0.4\\
30	0.4\\
31	1\\
32	1\\
};
\end{axis}
\end{tikzpicture}%
}
			\caption{}
			\label{fig:u_trace}
		\end{subfigure}
		\caption{\textit{Case study 4:} (a) \textsc{mc} for the discrete modes of the $CO_2$ model and (b) input control signal.}
	\end{minipage}
	\hfill 	
	\begin{minipage}{0.6\columnwidth}
		\begin{lstlisting}[caption={\textit{Case study 4:}  Definition of intial conditions for simulation},captionpos=b, label={lst:CO2:sim}]
		// Number of simulations
		int monte = 5000;
		// Initial continuous variables
		arma::mat x_init = arma::zeros<arma::mat>(2,monte);
		// Initialise random generators
		std::random_device rand_dev;
		std::mt19937 generator(rand_dev());
		// Define distributions
		std::normal_distribution<double> d1{450,25};
		std::normal_distribution<double> d2{17,2};
		for(size_t i = 0; i < monte; ++i)
		{
			x_init(0,i) = d1(generator);
			x_init(1,i) = d2(generator);
		}
		// Initial discrete mode q_0 = (E,C)
		arma::mat q_init = arma::zeros<arma::mat>(1,monte);
		// Definition of control signal
		// Read from .txt/.mat file or define here
		arma::mat u =readInputSignal("../u.txt");  
		//Combining
		exdata_t data(x_init,u,q_init);
		\end{lstlisting}
	\end{minipage}
\vspace{-0.65
	cm}
\end{figure}

For this last case study, we refer to the $CO_2$ model described in Case Study 1 (Sec.~\ref{CS1}). 
We extend the $CO_2$ model to capture (i) the effect of occupants leaving or entering the zone within a time step (ii) the opening or closing of the windows in the zone~\cite{Abate17memocode}. $\rho_m$ is now a control input and is an exogenous signal. 
This can be described as a \textsc{shs} comprising two-dimensional dynamics, over 
discrete modes in the set $\{q_0 = (E,C), q_1 = (F,C), q_2 = (F,O), q_3 =(E,O)\}$ 
describing possible configurations of the room  
(empty (E) or full (F), and with windows open (O) or closed (C)).  
A \textsc{mc} representing the discrete modes and their dynamics is in Figure~\ref{fig:SHS:CO2}. 
The continuous variables evolve according to Eqn.~\eqref{eqn:CO2}, 
which now captures the effect of switching between discrete modes, as  
\begin{align}\label{eqn:CO2:ex}
&x_{1,k+1} = x_{1,k} + \frac{\Delta}{V}(-\rho_mx_{1,k} + \varrho_{o,c}(C_{out} - x_{1,k})) + \mathbf{1}_{F}C_{occ,k}  + \sigma_{1} w_{k}, \\ 
&x_{2,k+1} = x_{2,k} + \frac{\Delta}{C_z}(\rho_mC_{pa} (T_{set} -x_{2,k}) +  \frac{\varrho_{o,c}}{R}(T_{out} - x_{2,k})) + \mathbf{1}_{F}T_{occ,k} + \sigma_{2} w_k, \nonumber 
\end{align}
where the additional terms are: 
$\varrho_{(\cdot)}$ is the natural drift air flow that changes depending whether the window is open ($\varrho_{o}$) or closed ($\varrho_{c}$) 
[$m^3/min$];  
$C_{occ}$ is the generated $CO_2$ level when the zone is occupied (it is multiplied by the indicator function $\mathbf{1}_{F}$)  [$ppm/min$];  
$T_{occ}$ is the generated heat due to occupants [$\oC/min$], which couples the dynamics in \eqref{eqn:CO2:ex} as $T_{occ,k} = vx_{1,k} + \hbar$.  

\paragraph{Implementation}
The provided file \textsc{cs4.mat}  
sets the values of the parameters in~\eqref{eqn:CO2:ex} and contains the transition probability matrix representing the relationships between discrete modes. We select a sampling time $\Delta = $ 15 [$min$] and  
simulate the evolution of this dynamical model over a fixed time horizon $K = 8$ hours (i.e. 32 steps)  
with an initial CO$_2$ level $x_1\sim \N(450,25)\; [ppm]$ and a temperature level of $x_2\sim \N(17,2) \;[\oC]$.
We define the  initial conditions using Listing~\ref{lst:CO2:sim}.
Line 2 defines the number of Monte Carlo simulations using by the variable \texttt{monte} and sets this to 5000.
We instantiate the initial values of the continuous variables using the term \texttt{x\_init}, while we set the initial discrete mode using the variable \texttt{q\_init}. This is done using  lines 4-17 which defines independent normal distribution for each of the continuous variable from which we sample 5000 points for each of the continuous variables and defines the initial discrete mode to $q_0 = (E,C)$. 
We define the control signal $\rho_m$ in line 20, by parsing the \texttt{u.txt} which contains discrete values of $\rho_m$ for each time step (see Fig.~\ref{fig:u_trace}). 
Once the model is defined, we follow Listing~\ref{lst:input} to perform the simulation. 
The simulation engine also generates a \python~script, \texttt{simPlots.py}, which gives the option to visualise the simulation outcomes offline. 

\begin{figure}[t]
	\centering
	\begin{subfigure}{.3\columnwidth}
		\centering
		\resizebox{\columnwidth}{!}{
%
%
\definecolor{mycolor1}{rgb}{0.00000,0.44700,0.74100}%
\begin{tikzpicture}

\begin{axis}[%
width=2.521in,
height=1.466in,
scale only axis,
xmin=0,
xmax=32,
xlabel style={font=\color{white!15!black}},
xlabel={\large Time steps},
ymin=370,
ymax=415,
ylabel style={font=\color{white!15!black}},
ylabel={\large $\text{x}_\text{1}$},
axis background/.style={fill=white},
legend style={legend cell align=left, align=left, draw=white!15!black}
]
\addplot [color=mycolor1]
  table[row sep=crcr]{
0	412.868462943865\\
1	413.446636771293\\
2	406.999729101338\\
3	400.678626387654\\
4	400.578473809851\\
5	400.479894829351\\
6	394.382866721792\\
7	388.381635119859\\
8	382.474694868856\\
9	382.660562463077\\
10	382.843509705657\\
11	383.023582481161\\
12	383.200825953218\\
13	377.375286575841\\
14	377.641276073161\\
15	377.903086343499\\
16	378.16078305087\\
17	378.414430827575\\
18	378.651141624406\\
19	378.897084887203\\
20	379.139163889455\\
21	379.377439346424\\
22	379.611971019413\\
23	379.842817730756\\
24	380.070037378571\\
25	380.274440669165\\
26	380.494878657371\\
27	374.711855123907\\
28	375.020817983019\\
29	375.32380089087\\
30	375.622023331272\\
31	375.913225167986\\
32	376.198744492058\\
};

\end{axis}
\end{tikzpicture}%
}
		\caption{}
		\label{fig:x1_trace}
	\end{subfigure}
	\begin{subfigure}{.3\columnwidth}
		\centering
		\resizebox{\columnwidth}{!}{
%
%
\definecolor{mycolor1}{rgb}{0.00000,0.44700,0.74100}%
\begin{tikzpicture}

\begin{axis}[%
width=2.521in,
height=1.566in,
scale only axis,
xmin=0,
xmax=32,
xlabel style={font=\color{white!15!black}},
xlabel={ \large Time steps},
ymin=17.9,
ymax=20,
ylabel style={font=\color{white!15!black}},
ylabel={\large $\text{x}_\text{2}$},
axis background/.style={fill=white},
legend style={legend cell align=left, align=left, draw=white!15!black}
]
\addplot [color=mycolor1]
  table[row sep=crcr]{%
0   18.2\\
1	18.2063020572976\\
2	18.061386556979\\
3	17.9249279185574\\
4	18.0424698327805\\
5	18.1543752531386\\
6	18.2227818116349\\
7	18.2877489234889\\
8	18.3494439680342\\
9	18.4461588323461\\
10	18.5382429934814\\
11	18.6259183268598\\
12	18.709396074488\\
13	18.7507447846071\\
14	18.8280906717238\\
15	18.901736333871\\
16	18.9718590149133\\
17	19.0386274650414\\
18	19.1597412078842\\
19	19.2175172745366\\
20	19.2725305811637\\
21	19.3249134466735\\
22	19.3747918494497\\
23	19.4222857312311\\
24	19.4675092864274\\
25	19.5631354973542\\
26	19.6016194136117\\
27	19.6001320120294\\
28	19.6877220372224\\
29	19.720095993476\\
30	19.7509262313026\\
31	19.82956352324\\
32	19.903527242889\\
};

\end{axis}
\end{tikzpicture}%
}
		\caption{} 
		\label{fig:x2_trace}
	\end{subfigure}
	\begin{subfigure}{.3\columnwidth}
		\centering
		\resizebox{\columnwidth}{!}{
%
%
\definecolor{mycolor1}{rgb}{0.00000,0.44700,0.74100}%
\begin{tikzpicture}

\begin{axis}[%
width=2.521in,
height=1.566in,
scale only axis,
xmin=0,
xmax=32,
xlabel style={font=\color{white!15!black}},
xlabel={\large Time steps},
ymin=0,
ymax=3,
ytick={0,1,2,3},
yticklabels={\normalsize {(E,C)},{\normalsize(F,C)},{\normalsize(F,O)},{\normalsize(E,O)}},
ylabel style={font=\color{white!15!black}},
axis background/.style={fill=white},
legend style={legend cell align=left, align=left, draw=white!15!black}
]
\addplot[const plot, color=mycolor1] table[row sep=crcr] {%
0	0\\
1	0\\
2	2\\
3	2\\
4	2\\
5	1\\
6	2\\
7	3\\
8	1\\
9	3\\
10	2\\
11	2\\
12	3\\
13	1\\
14	1\\
15	2\\
16	2\\
17	2\\
18	1\\
19	2\\
20	2\\
21	2\\
22	3\\
23	2\\
24	1\\
25	2\\
26	2\\
27	2\\
28	1\\
29	2\\
30	2\\
31	3\\
32	2\\
};
\end{axis}
\end{tikzpicture}%
}
		\caption{} 
		\label{fig:q_trace}
	\end{subfigure}
	\\ 
	\begin{subfigure}{.3\columnwidth}
		\centering
		\includegraphics[width=\columnwidth]{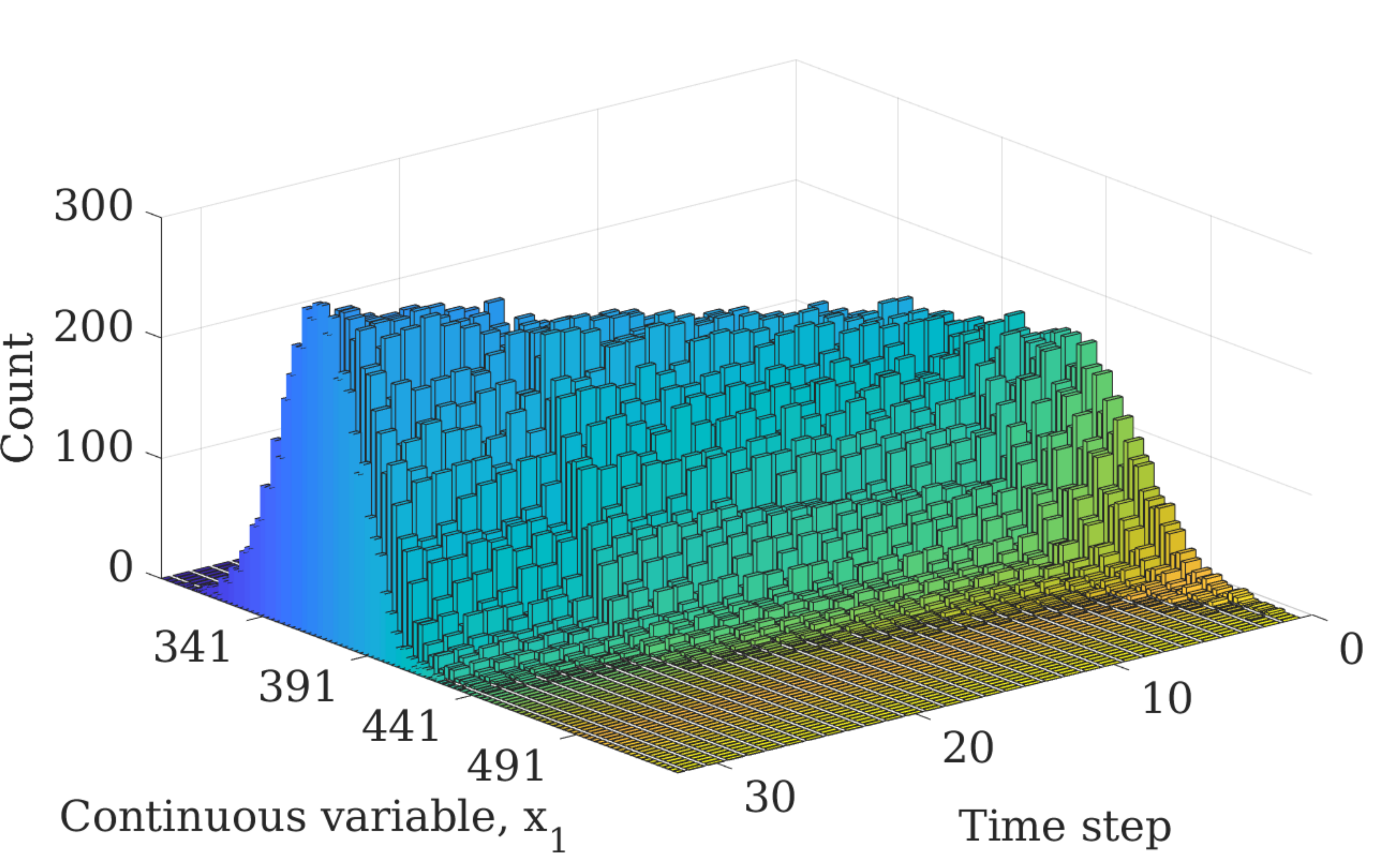}
		\caption{}
		\label{fig:x1_hist}
	\end{subfigure}
	\begin{subfigure}{.3\columnwidth}
		\centering
		\includegraphics[width=1\columnwidth]{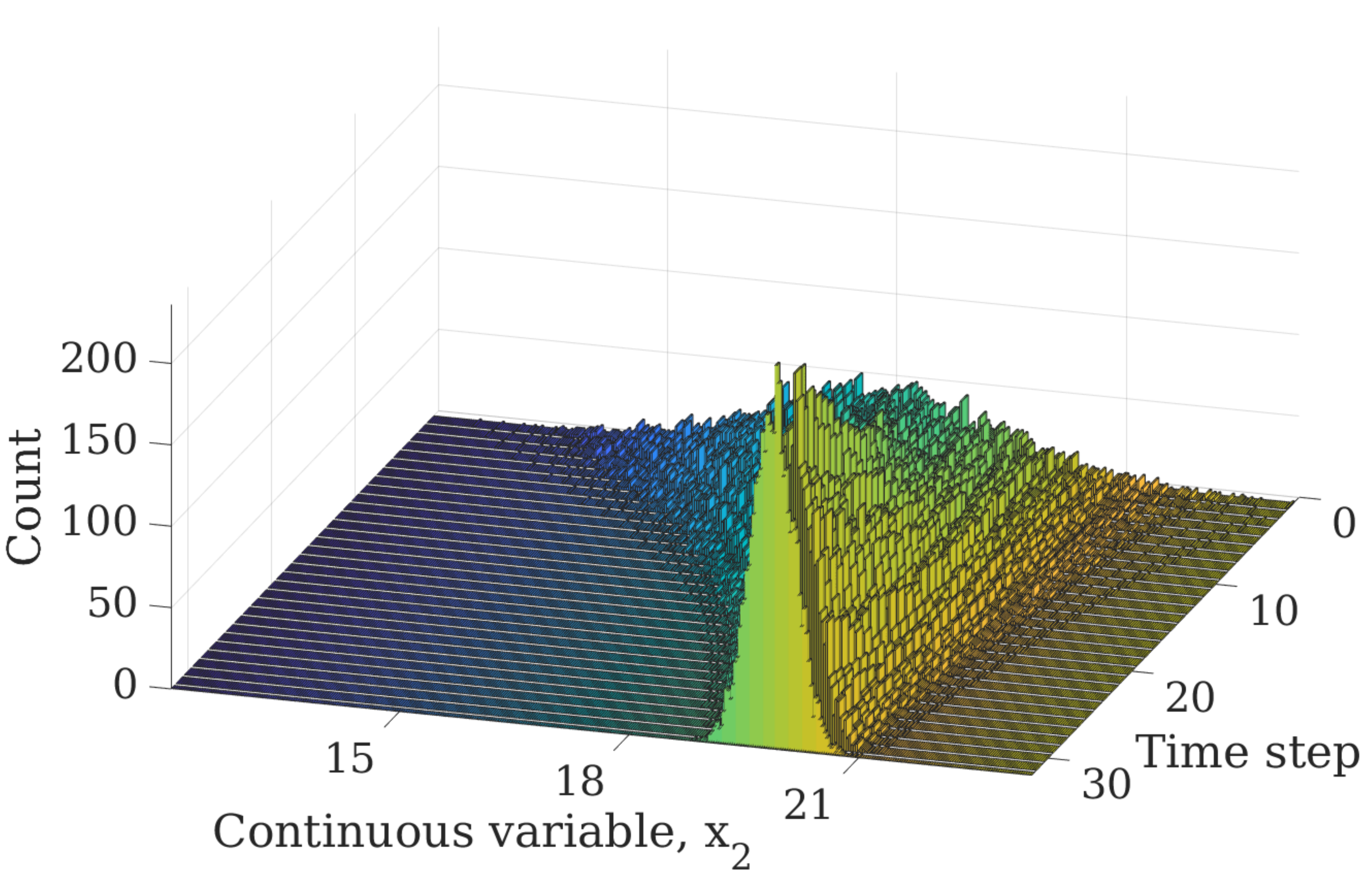}
		\caption{} 
		\label{fig:x2_hist}
	\end{subfigure}
	\begin{subfigure}{.3\columnwidth}
		\centering
		\includegraphics[width=\columnwidth]{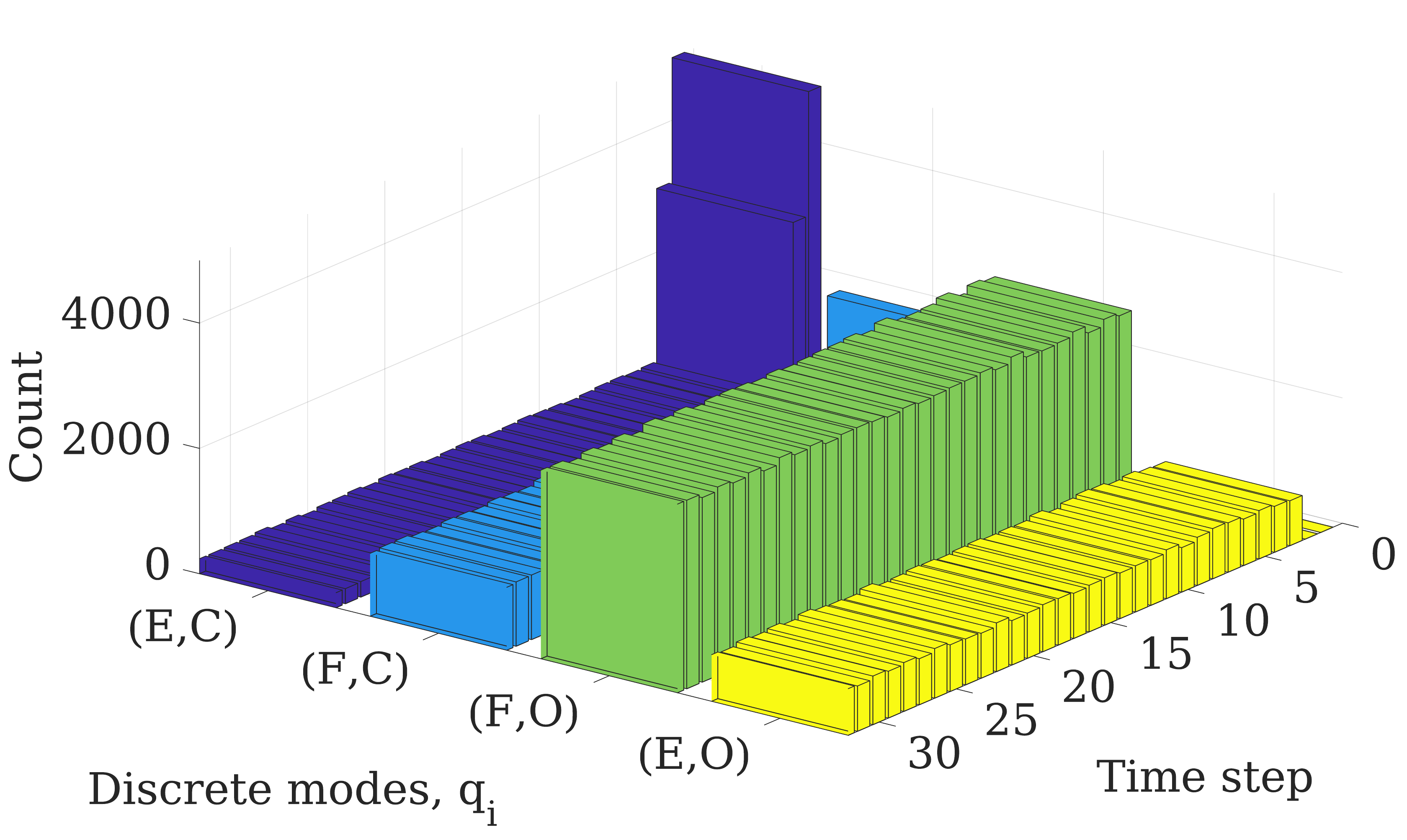}
		\caption{} 
		\label{fig:q_hist}
	\end{subfigure}
	\caption{\textit{Case study 4:}  Simulation single traces for continuous variables (a) $x_1$, (b) $x_2$ and discrete modes (c) $q$. Histogram plots with respect to time step for (d) $x_1$, (e) $x_2$ and discrete modes (f) $q$. }
	\label{fig:Cs4:sim}
	\vspace{-.5cm}
\end{figure}

\paragraph{Outcomes}
The generated simulation plots are shown in Fig.~\ref{fig:Cs4:sim}, which depicts:
(i) a sample trace for each continuous variable (the evolution of $x_1$ is shown in Fig.~\ref{fig:x1_trace}, $x_2$ in Fig.~\ref{fig:x2_trace}) and for the discrete modes (see Fig.~\ref{fig:q_trace}); and 
(ii) histograms depicting the range of values the continuous variables can be in during each time step and the associated count (see Fig.~\ref{fig:x1_hist} for $x_1$ and Fig.~\ref{fig:x2_hist} for $x_2$); and a histogram showing the likelihood of being in a  discrete mode within each time step (see Fig.~\ref{fig:q_hist}). The total time taken to generate the simulations is 48.6 [s].


\section{Conclusions and Extensions} 
We have presented \stochy, a new software tool for the quantitative analysis of stochastic hybrid systems. 
There is a plethora of enticing extensions that we are planning to explore. In the short term, we intend to: 
(i) interface with other model checking tools such as \textsc{storm}~\cite{storm} and the \modest~\cite{hahn2013compositional}; 
(ii) embed algorithms for policy refinement, so we can generate policies for models having numerous continuous input variables~\cite{peva}. 
In the longer term, we plan to extend \stochy~such that 
(i) it employs a graphical user-interface; 
(i) it may allow analysis of continuous-time \shs;and 
(iii) it makes use of data structures such as multi-terminal binary decision diagrams~\cite{Fujita1997} to reduce the memory requirements during the construction of the abstract \mdp~or \imdp.

\paragraph{Acknowledgements}
The author's would also like to thank Sadegh Soudjani, Sofie Haesaert, Luca Laurenti, Morteza Lahijanian and Viraj Brian Wijesuriya.
This work is in part funded by the Alan Turing Institute, UK, and by Malta's ENDEAVOUR Scholarships Scheme.

%
%
%
\bibliographystyle{splncs04}
\bibliography{Bib_rep}

\end{document}